\documentclass[preprint,showpacs,eqsecnum,aps]{revtex4}
\usepackage{graphicx}
\usepackage{epsfig}

\begin{document}

\title{Phenomenological description of the states $0^+$ and $2^+$ in some even-even nuclei}

\author{A. A. Raduta$^{a),b),c),d)}$,  F. D. Aaron $^{a)}$, E. Moya de Guerra $^{d), e)}$and Amand Faessler$^{c)}$ }

\address{$^{a)}$ Department of Theoretical Physics and Mathematics, Bucharest
  University, POBox MG11, Romania}

\address{$^{b)}$ Department of Theoretical Physics, Institute of Physics and
  Nuclear Engineering, Bucharest, POBox MG6, Romania}

\address{$^{c)}$ Institut fuer Theoretische Physik der Universitaet Tuebingen, Auf der Morgenstelle 14, Germany}
 
\address{$^{d)}$ Departamento de Fisica Atomica, Molecular y Nuclear, Universidad Complutense de Madrid, E-28040 Madrid, Spain}
 
\address{$^{e)}$ Instituto de Estructura de la Materia, CSIC, Serrano 123, E-28006 Madrid, Spain}

\begin{abstract}A sixth-order quadrupole boson Hamiltonian is used to describe the states $0^+$ and $2^+$ identified in several nuclei by various types of experiments. Two alternative descriptions of energy levels are proposed. One corresponds to a semi-classical approach of the model Hamiltonian while the other one provides the  exact eigenvalues. Both procedures yield close formulas for energies. The first procedure involves four parameters, while the second involves a compact formula with five parameters. In each case the  parameters are fixed by a least-square fit procedure. Applications are performed for eight even-even nuclei.
 Both methods yield results which are in a surprisingly good agreement with the experimental data. We give also our predicted  reduced transition probabilities within  the two approaches, although the corresponding experimental data are
 not yet available.
\end{abstract}
\maketitle


\renewcommand{\theequation}{1.\arabic{equation}}
\setcounter{equation}{0}
\section{Introduction}
\label{sec:level1}

The collective states of deformed nuclei are usually classified in rotational bands distinguished by a quantum number K, which is the angular momentum projection on the $z$ axis of the intrinsic reference frame. The collective character of the states is diminished by increasing the value of K \cite{Bohr,Grei1,Grei2,Ring}. In Ref. \cite{Rad1} one of us (A.A.R.) suggested a possible method of developing bands in a {\it horizontal} fashion.
Indeed, therein on the top of each state in the ground band a full band of monopole multi-phonon states has been constructed. The states of the newly constructed band on the top of the ground band state of angular momentum $J$ have the same angular momentum $J$. This feature contrasts the property of the ground band, where the states have different angular momenta, i.e. 0, 2, 4, 6 etc. 
This idea has been recently considered in a phenomenological context trying to organize the states, describing the motion of the intrinsic degrees of freedom, in bands. Thus, two intrinsic collective coordinates, similar to the nuclear deformations $\beta$ and $\gamma$, are described by the irreducible representations of a SU(2) group acting in a fictitious space (i.e. not in ordinary space) . Compact formulas for the excitation energies have been obtained \cite{Rad2,Rad3}.

Recently, about 26 states $0^+$ and 67 states $2^+$ have been populated in $^{168}Er$ by means of a $(p,t)$ reaction \cite{Bucur}. In the cited paper the excitation energies and the corresponding reaction strength have been provided. These data were described qualitatively by two microscopic models, called projected shell model (PSM) and quasiparticle phonon model (QPM), respectively.
Both models have some inherent drawbacks. PSM restricts the fermion space to four quasiparticle states and even from the four qp space the states with four alike quasiparticles are excluded. This is not the case of QPM where the multi-quasiparticle components are taken into account by means of the QRPA approach. However, the final states contain at most two phonon states. These states  violate the Pauli principle and moreover are not states of good angular momentum.

In Ref.[9] some of us made a first attempt to fit the data of Ref.[8] using a phenomenological model, namely a sixth-order quadrupole boson Hamiltonian that was developed in Ref. [7].
Since then about 12 new $2^+$ states have been identified by a more careful analysis of the data produced in the $(p,t)$ experiment [10].
Here we show that the  complete $0^+$ and $2^+$ data sets, presently available, are nicely described by the closed formulas provided by the model of 
Ref. [7].

Here we present details about both the semi-classical approach and the boson description of these states. The compact analytical formulas are used to explain the data about the states $0^+$ and $2^+$ in several even-even nuclei:   $^{152,154}$Gd, $^{162}$Dy, $^{168}$Er, $^{176}$Hf, $^{180,184}$W, $^{190}$Os.
The model quadrupole boson Hamiltonian is presented in Section 2. Therein we also present two distinct approaches for its spectrum.
Analytical formulas for  the reduced transition probabilities, corresponding to the mentioned treatments, are derived in Section III. Numerical applications to eight nuclei are presented in Section IV. The final conclusions are summarized in Section V. 

\section{The model Hamiltonian}
\label{sec:level2}
\renewcommand{\theequation}{2.\arabic{equation}}
\setcounter{equation}{0}

We attempt to describe the set of states $0^+$ and $2^+$ identified in various experiments in terms of quadrupole bosons, by means of the model Hamiltonian: 
\begin{equation}
H=\epsilon \hat{N}+
\sum_{J=0,2,4}C_J\left[\left(b^{\dagger}_2b^{\dagger}_2\right)_J\left(b_2b_2\right)_J\right]_0
+F\left(b^{\dagger}_2b^{\dagger}_2\right)_0\hat{N}\left(b_2b_2\right)_0 ,
\label{hahmodel}
\end{equation}
where $b^{\dagger}_{2\mu}, b_{2\mu}$, with $-2\leq \mu \leq 2$, are the quadrupole boson operators and $\hat{N}$ the boson number operator.
The first remark about the chosen Hamiltonian refers to the fact that it commutes with the boson number operator.We recall that this feature
is one of the signatures of the interacting boson approximation (IBA)
\cite{AriIache}
which, as a matter of fact, was very successful in describing rotational bands in non-spherical nuclei. Moreover, the Hamiltonian given by Eq. (\ref{hahmodel}) with $F=0$, has been used, even before the IBA was proposed,
to describe the yrast bands in transitional and deformed nuclei \cite{RadDreiz,DreizKlei}. Thus, an analytical formula for the yrast energies has been obtained, which in fact was generalizing the empirical expression used by Ejiri \cite{Ejiri}.
As in Ref.\cite{Rad3}, this Hamiltonian is alternatively  treated semi-classically and exactly solved in the boson space.
For a self-contained presentation we give here the basic results obtained in the mentioned treatments.

\subsection{Semi-classical treatment}
The boson Hamiltonian (\ref{hahmodel}) is treated by a Time Dependent Variational Principle (TDVP):
\begin{equation}
\delta \int_{0}^{t}\langle \Psi|\left(H-i\hbar\frac{\partial}{\partial t'}\right)
|\Psi\rangle dt'=0.
\label{delta}
\end{equation}
When the variational state $|\Psi\rangle$ spans the whole space of the boson states, solving the equation
(\ref{delta}) is equivalent to solving the time dependent equation associated to the
model Hamiltonian $H$.
The classical features encountered by $H$ can be described by restricting the space of
$|\Psi\rangle$ to the coherent states:
\begin{equation}
|\Psi\rangle = \rm{exp}\left[z_0b^{\dagger}_0-z^*_0b_0+z_2(b^{\dagger}_{2}+
b^{\dagger}_{-2})-
z^*_2(b_{2}+b_{-2})\right]|0\rangle.
\label{Psi}
\end{equation}
Here the boson vacuum state is denoted by $|0\rangle$. The function $|\Psi\rangle$ depends on the complex parameters $z_0, z_2$ and their complex conjugates $z_0^*,z_2^*$.  These parameters play the role of classical phase space coordinates whose equations of motion
are provided by the TDVP equations.
By a suitable change of coordinates,

\begin{eqnarray}
q_i&=&2^{(k+2)/4}Re(z_k),\;p_i=\hbar2^{(k+2)/4}Im(z_k),\nonumber\\
k&=&0,2,\;i=\frac{k+2}{2},
\end{eqnarray}
the classical equations of motion acquire a canonical form, while the classical Hamilton function (the average of $H$ with $|\Psi\rangle$), ${\cal H}$, becomes a function of the generalized phase space coordinates, $q$ and $p$: 

\begin{eqnarray}
{\cal H}&=&\frac{A}{2}\left(q_1^2+q_2^2+\frac{1}{\hbar^2}(p_1^2+p_2^2)\right)+
\frac{B}{4}\left(q_1^2+q_2^2+\frac{1}{\hbar^2}(p_1^2+p_2^2)\right)^2+\frac{C}{8\hbar^2}(q_1p_2-q_2p_1)^2
\nonumber\\
&+&
\frac{F}{10}\left[\frac{1}{4}\left(q_1^2+q_2^2+\frac{1}{\hbar^2}(p_1^2+p_2^2)\right)^2-
\frac{1}{\hbar^2}(q_1p_2-q_2p_1)^2\right]\left(q_1^2+q_2^2+\frac{1}{\hbar^2}(p_1^2+p_2^2)\right).\nonumber\\
\label{Hrond2}
\end{eqnarray}
where the factors $A$, $B$ and $C$ have simple expressions in terms of the coefficients
$\epsilon, C_J$ involved in the boson Hamiltonian:
\begin{eqnarray}
A&=&\epsilon,\; B=\frac{1}{5}C_0+\frac{2}{7\sqrt{5}}C_2+\frac{6}{35}C_4,
\nonumber\\
C&=&-\frac{8}{5}C_0+\frac{16}{7\sqrt{5}}C_2-\frac{8}{35}C_4.
\end{eqnarray}

Conventionally, we shall call the part of ${\cal H}$ not depending on momenta, as the potential energy of the system:
\begin{equation}
V(q)={\cal H}|_{p_1=p_2=0},~~\rm{with}~~ (q)=(q_1,q_2).
\end{equation}
Thus, the potential energy associated to ${\cal H}$ is:
\begin{equation}
V(q)=\frac{A}{2}\left(q_1^2+q_2^2\right) +\frac{B}{4}\left(q_1^2+q_2^2\right)^2+
\frac{F}{40}\left(q_1^2+q_2^2\right)^3.
\label{Vdeq}
\end{equation}

In Ref.\cite{Rad3} we showed that this classical function exhibits a symmetry with respect to the classical rotations generated by the classical functions obtained by averaging the  generators of a $SU_b(2)$ algebra with $|\Psi\rangle $:

\begin{eqnarray}
\hat{L}_1&=&\frac{\hbar}{4}\left[2b^{\dagger}_0b_0-(b^{\dagger}_2+b^{\dagger}_{-2})
(b_2+b_{-2})\right],\nonumber\\
\hat{L}_2&=&\frac{\hbar}{2\sqrt{2}}\left[b^{\dagger}_0(b_2+b_{-2})
+(b^{\dagger}_2+b^{\dagger}_{-2})b_0\right],\nonumber\\
\hat{L}_3&=&\frac{\hbar}{2\sqrt{2}i}\left[b^{\dagger}_0(b_2+b_{-2})
-(b^{\dagger}_2+b^{\dagger}_{-2})b_0\right].
\label{eloper}
\end{eqnarray}

\noindent
Thus, the generators of the classical $SU_c(2)$ algebra acting in a fictitious space are defined by:
\begin{equation}
{\cal L}_k=\langle \Psi |{\hat L}_k|\Psi \rangle, k=1,2,3.
\end{equation}
It can be checked that the classical system has two constants of motion and these are ${\cal H}$ and ${\cal L}_3$.
On the other hand the system is fully described by two degrees of freedom, $q_1$ and $q_2$. Consequently, the classical system is fully solvable (or integrable).Therefore, the equations of motion can be integrated and the trajectories analytically described.

 ${\cal H}$ contains two distinct terms describing an anharmonic motion of a classical plane oscillator and a pseudo-rotation around an axis perpendicular to the oscillator plane, respectively. Taking into account that 
the third component of the pseudo-angular momentum is a constant of motion, the classical Hamiltonian considered in the reduced space can be easily quantized and the resulting energy is:
\begin{equation}
\epsilon_{n,M}=A(n+1)+B(n+1)^2+\frac{C}{2}M^2
+\frac{F}{5}\left[(n+1)^3-4(n+1)M^2\right],
\label{eclass}
\end{equation}   

The number of the  oscillator quanta in the $q_1,q_2$ plane is denoted by $n$ while the value of the third component of the pseudo-angular momentum is $M$. 
Actually, Eq. (\ref{eclass}) represents a semi-classical spectrum which describes the motion of the intrinsic degrees of freedom $q_1$ and $q_2$, related to the nuclear deformations $\beta$ and $\gamma$. 

Assuming that the rotational degrees of freedom are only weakly coupled to the motion of the intrinsic coordinates, the total energy associated to the motion in the laboratory frame can be written as a sum of two terms corresponding to the intrinsic and rotational motions, respectively: 
\begin{eqnarray}
\epsilon_{n,M,J}&=&A(n+1)+B(n+1)^2+\frac{C}{2}M^2
\nonumber\\
&+&\frac{F}{5}\left[(n+1)^3-4(n+1)M^2\right]+\delta J(J+1)
\label{enmj}
\end{eqnarray}
Averaging both angular momenta squared, ${\hat J}^2$ and ${\hat L}^2$, on $|\Psi\rangle$ one obtains a relationship between the two momenta.
Thus (see Ref.\cite{Rad3}), to the values $J=0$ and $J=2$ correspond different values of $M$, namely $M=0$ and $M=1$,
 respectively. 

Therefore, considering the above equation for the sets of states with angular momenta J=0, 2 and normalizing the results to the energy of the first $0^+$ state, one obtains the following expressions for the excitation energies:
\begin{eqnarray}
E_{n,0}&\equiv&E_{n,0,0}-E_{0,0,0}=
\frac{1}{5}Fn^{3}+(\frac{3}{5}F+B)n^{2}+(A+2B+\frac{3}{5}F)n,\;\; n\geq 0,\nonumber\\
E_{n,1}&\equiv &E_{n,1,2}-E_{0,0,0}=  \frac{1}{5}Fn^{3}+(\frac{3}{5}F+B)n^{2}+(A+2B-\frac{1}{5}F)n+{\cal C},\;\; n\geq 1,
\label{En01}
\end{eqnarray}
where
\begin{equation}
{\cal C}=-\frac{4}{5}C_0+\frac{2}{7\sqrt{5}}C_2+\frac{6}{35}C_4-\frac{4}{5}F .
\end{equation}

\subsection{Exact eigenvalues}

Note that the model Hamiltonian is highly anharmonic due to the terms of fourth and sixth-order in the quadrupole phenomenological bosons. Despite  this fact it is easy to see that this Hamiltonian is diagonal in the boson basis $|Nv\alpha JM\rangle $, where the quantum numbers have the significance of the boson number ($N$), seniority ($v$), missing quantum number ($\alpha$), angular momentum ($J$) and its projection on the axis OZ ($M$). These basis states have been analytically studied in Ref.\cite{Ghe} using alternatively different representations like, laboratory frame coordinates, intrinsic frame coordinates, boson variables. To prove the  statement concerning the diagonal form of H in the mentioned boson basis, it is useful to write the fourth order term in a different form
(see Ref. \cite{RadDreiz}) which results in having a more convenient expression for $H$:
\begin{eqnarray}
H&=&(A+\gamma )\hat{N}+(B+\frac{C}{8})\hat{N}^2-\frac{1}{6}\left(B+\frac{C}{8}+\gamma\right)\hat{J}^2\nonumber\\
&-&\frac{5}{8}C\left(b^{\dagger}_2b^{\dagger}_2\right)_0\left(b_2b_2\right)_0+F\left(b^{\dagger}_2b^{\dagger}_2\right)_0\hat{N}\left(b_2b_2\right)_0,
\end{eqnarray}
where the coefficient $\gamma$ has the expression:
\begin{equation}
\gamma =\frac{2}{7\sqrt{5}}C_2-\frac{3}{7}C_4.
\label{bosh}
\end{equation}
From this expression it is obvious the $H$ commutes with the operators ${\hat N}$, ${\hat \Lambda}$, ${\hat J}^2$, ${\hat J}_z$
where ${\hat \Lambda}$ denotes the Casimir operator of the group $R_5$:
\begin{equation}
\hat{\Lambda} ={\hat N}({\hat N}+3)-5(b^{\dagger}_2b^{\dagger}_2)_0(b_2b_2)_0.
\end{equation}
The eigenvalue corresponding to the state $|Nv\alpha JM\rangle $ is:
\begin{eqnarray}
&&E_{N,v,J}=\frac{1}{5}FN^{3}+(B+\frac{1}{5}F)N^{2}\\
&+& 
(A+\gamma -3(\frac{1}{8}C+\frac{2}{5}F))N-\frac{1}{6}(B+\frac{1}{8}
C+\gamma )J(J+1)\nonumber\\
&+&(\frac{1}{8}C+\frac{2}{5}F)v^{2}+3(\frac{1}{8}C+
\frac{2}{5}F)v
-\frac{1}{5}FNv^{2}-\frac{3}{5}FNv.\nonumber 
\label{envj}
\end{eqnarray}
Comparing this with Eq.(\ref{enmj}), we notice that the eigenvalues of $H$, corresponding to a given $J$, are characterized by two quantum numbers, namely the number of bosons $N$ and the seniority $v$.
Therefore, using the new expression for energies one expects a better description of the data.
For $J = 0$ we  use the lowest two values for seniority quantum number, i.e. $v=0,3$,  and obtain: 
\begin{eqnarray}
E_{N,0,0}&=&\frac{1}{5}FN^{3}+(B+\frac{1}{5}F)N^{2}+ 
(A+\gamma -\frac{3}{8}C-\frac{6}{5}F)N,\;N=0,2,4,...
\label{jeq0}\\
E_{N,3,0}&=&\frac{1}{5}FN^{3}+(B+\frac{1}{5}F)N^{2}+ 
(A+\gamma -\frac{3}{8}C-\frac{24}{5}F)N+\frac{9}{4}C+
\frac{36}{5}F, N=3,5,7,....\nonumber
\end{eqnarray}
Similarly, for $J=2$ we consider the lowest two allowed seniorities, i.e. $v=1,2$. The result is:
\begin{eqnarray}
E_{N,1,2}&=&\frac{1}{5}FN^{3}+(B+\frac{1}{5}F)N^{2}+ 
(A+\gamma -\frac{3}{8}C-2F)N-B-\gamma +\frac{3}{8}C+
\frac{8}{5}F, N=1,3,5,... \nonumber\\
E_{N,2,2}&=&\frac{1}{5}FN^{3}+(B+\frac{1}{5}F)N^{2}+ 
(A+\gamma -\frac{3}{8}C-\frac{16}{5}F)N-B-\gamma +\frac{9}{8}
C+4F, N=2,4,6,....\nonumber\\
\label{jeq2}
\end{eqnarray}
Note that within the boson treatment the lowest energy denoted by $E_{0,0,0}$ is equal to zero and therefore there is no need to renormalize the energies of excited states with respect to the ground state energy.
Let us now turn our attention to the missing quantum number $\alpha$. This quantum number labels the $R_3$ irreducible representations ($J$) which appear in an $R_5$ irreducible representation ($v$). The name is suggesting that there is no intermediate group between $R_5$ and $R_3$ whose Casimir operator might make the distinction between different $J$ representations corresponding to the same seniority $v$.
 $\alpha$ labels the solutions of the double inequality for the integer number $p$ \cite{GheA}: 
\begin{equation}
v-J\leq 3p\leq v-\frac{J}{2}, {\rm{for}}\; J={\rm even},\; p={\rm integer}.
\label{ineq}
\end{equation}
The number of solutions for this double inequality is the degeneracy $d_v(I)$, characterizing the reduction $R_5\supset R_3$.
It is clear that for $J=2$ and fixed  $v$, the number of solutions of Eq. (2.21) is either 0 or 1. For example, there is no state $2^+$ with $v=0,3,6,9,...$.
Concerning the states  $J=0$, one has $d_v(0)=1$ if $v=3k$ with $k$ positive integer and $d_v(0)=0$ otherwise.

Concluding, for $J=0,2$ there is no degeneracy, i.e. the set $(v,J)$ either does not exist or is uniquely determined by
the relation (\ref{ineq}).

\section{Electric quadrupole transitions}
\label{sec:level3}
\renewcommand{\theequation}{3.\arabic{equation}}
\setcounter{equation}{0}

The states $0^+$ can be related to the states $2^+$ by E2 transitions whereas the states of the same angular momentum are related by E0 transitions.
 Since the E0 transitions for highly excited states are not yet  experimentally investigated we confine our study to the E2 transitions.
Compact formulas for E2 transitions have been presented in our previous publication \cite{Rad07}. However details about the derivation of these expressions were not given. Here we complete the description of the E2 transitions by providing additional information which will facilitate a straightforward derivation of the results listed in the reference quoted above.
\subsection{Semi-classical approach}
We suppose that the leading contribution to the E2 transitions is provided by the linear boson term:
\begin{equation}
T_{2\mu}=q_h\left(b^{\dagger}_{2\mu}+(-)^{\mu}b_{2,-\mu}\right).
\end{equation}
The average of this operator with the coherent state $|\Psi\rangle$ (see Eq.(\ref{Psi}))has the expression:
\begin{equation}
\langle \Psi | T_{2\mu}|\Psi \rangle \equiv 
{\cal  T}_{2\mu}=q_{h}\left[\delta_{\mu,0}\sqrt{2}q_1+\left(\delta_{\mu,2}+\delta_{\mu,-2}\right)q_2\right].
\end{equation}
Here $\delta_{m,n}$ stands for the Kronecker symbol.

The semi-classical energies have been obtained by quantizing the plane oscillator defined with the coordinates $q_1$ and $q_2$. Thus, the energies depend on the total number of quanta along the two plane axes. It is convenient to use the polar coordinates associated to the Cartesian $q_1$ and $q_2$. The principal and radial quantum numbers are related by:
\begin{equation}
2n_r+\delta_{J,2}=n.
\end{equation}
Since the M-quantum number is equal to 0 for $0^+$ states and 1 for the states $2^+$, one can use only one label for the intrinsic states $|n_r,M\rangle$:

\begin{equation}
|0_n\rangle =|\frac{n}{2},0\rangle,~~|2_n\rangle =|\frac{n-1}{2},1\rangle .
\end{equation}
Using the explicit wave functions for the plane oscillator one calculates the matrix elements of the 
function ${\cal T}_{2\mu}$.
In the laboratory frame, the transition operator is acting on both the coordinates $q_1, q_2$ and the Euler angles $\Omega=(\theta_1,\theta_2,\theta_3)$
and has the expression:
\begin{equation}
{\bf T}_{2M}={q_h}\sqrt{2}\left(q_1 D^2_{M0}+\frac{q_2}{\sqrt{2}}(D^2_{M2}+D^2_{M,-2})\right),
\end{equation}
where $D^J_{MK}$ is the Wigner function describing the rotation matrix. 

In the liquid drop model the state of angular momentum 2 in the laboratory frame consists of two factors, one depending only on the deformation $\beta$, while the other one is linear combination of the Wigner functions $D^2_{MK}$ with the coefficients $g_K$ depending on the deformation $\gamma$.
In the present formalism by averaging the boson Hamiltonian on the coherent state $|\Psi\rangle $ one obtains the equations of motion for the intrinsic variable
$q_1, q_2$ which may be related to the deformations $\beta,\gamma$. Therefore, we assume that in the laboratory frame the wave functions are factorized in the following manner:
\begin{eqnarray}
|2nM\rangle &=&|2_n\rangle \Psi_{2M}(\Omega),~~\Psi_{2M}(\Omega)=\sqrt{\frac{5}{6}}\frac{1}{2\pi}\left[D^2_{M0}+D^2_{M2}+D^2_{M,-2}\right],
\nonumber\\
|0n0\rangle &=&|0_n\rangle \frac{1}{2\pi\sqrt{2}}.
\end{eqnarray}
Using the convention of Rose \cite{Rose} for the reduced matrix elements, we have:
\begin{equation}
B(E2;2^+_{n^{\prime}}\to
0^{+}_{n})\equiv \langle 2n'||{\bf T}_2||0n\rangle^2 .
\end{equation}
with $n\geq 1$. 
Analytical expressions for the above B(E2) values as well as for some particular branching ratios were given in Ref.\cite{Rad07}.

For a transition operator having an harmonic structure, the E2 transition
between any two states $2^+$ is forbidden.
This result is specific to the present semi-classical description. Indeed, the matrix elements of the variables $q_1$ and $q_2$ between the
states $|2_n\rangle $  and $|2_{n'}\rangle$ are equal to zero due to the integration over the polar angle. In Ref.\cite{Rad3} we gave a group theory argument for this result. Indeed, with respect to the pseudo-rotation group the harmonic transition operator is a tensor of rank $1/2$ while the states $2^+$  have the pseudo-angular momentum equal to 1. Then, it becomes manifest that two states $2^+$ cannot be linked by an harmonic transition operator. Of course, that is not true in the boson treatment, as we shall see in the next subsection.
 
In order to get non-vanishing transition matrix elements between two different $2^+$ states we introduced an anharmonic term in the expression of the transition operator:
\begin{equation}
T^{anh}_{2\mu}=q_{anh}\left[(q_1^2+q_2^2)D^2_{M0}+\frac{q_1q_2}{\sqrt{2}}(D^2_{M2}+D^2_{M,-2})\right].
\end{equation}

The reduced matrix element between two $2^+$ states can be analytically obtained \cite{Rad07}
A peculiar feature of the present formalism is the fact that
 the anharmonic term does not contribute to the transition $2^+\to 0^+$. On the other hand, as we have already mentioned, the harmonic term does not
contribute to the transition $2^+_n\to 2^+_{n'}$. Thus, the final result for the transition $2^+_n\to 2^+_{n-2}$ is
proportional to $(n-1)^2$.

\subsection{E2 transitions within the boson picture}

In what follows we shall identify the missing quantum number with the integer positive number $p$ which satisfies the inequality (\ref{ineq}). In the intrinsic frame of reference, the states $|n\lambda p I M\rangle $ have a factorized form \cite{Ghe,Apo}:  
\begin{equation}
|n\lambda p I M\rangle =F_{n\lambda}(\beta){\cal G}_{\lambda pI}(\gamma,\Omega),
\end{equation}

where

\begin{equation}
F_{n\lambda}(\beta)=\left[2(\frac{1}{2}(n-\lambda))!\right]^{\frac{1}{2}}\left[\Gamma(\frac{1}{2}(n+\lambda+5))\right]^{-\frac{1}{2}}
\beta^{\lambda}L^{\lambda+\frac{3}{2}}_{\frac{1}{2}(n-\lambda)}(\beta^2)exp(-\frac{1}{2}\beta^2).
\end{equation}
$L^{\lambda+\frac{3}{2}}_{\frac{1}{2}(n-\lambda)}(\beta^2)$ stands for the generalized Laguerre
polynomial. The functions $F_{n\lambda}$ are orthonormalized on the interval $[0,\infty)$ with the integration measure $\beta^4d\beta $. We need the normalized functions depending on the variables $\gamma$ and $\Omega$ for the angular momenta $0$ and $2$. These are \cite{Ghe,Apo}:
\begin{eqnarray}
{\cal G}_{000}(\gamma ,\Omega)&=&\frac{1}{4\pi\sqrt{2}},~~{\cal G}_{310}(\gamma,\Omega)=
\frac{1}{4\pi}\sqrt{\frac{3}{2}}\cos 3\gamma,\nonumber\\
{\cal G}_{102}(\gamma ,\Omega)&=&\frac{1}{4\pi}\sqrt{\frac{5}{2}}\left[\cos\gamma D^2_{M0}+\frac{\sin\gamma}{\sqrt{2}}\left(D^2_{M2}+D^2_{M,-2}\right)\right],\nonumber\\ 
{\cal G}_{202}(\gamma ,\Omega)&=&\frac{1}{4\pi}\sqrt{\frac{5}{2}}\left[\cos 2\gamma D^2_{M0}-\frac{\sin 2\gamma}{\sqrt{2}}\left(D^2_{M2}+D^2_{M,-2}\right)\right].
\end{eqnarray}
In the intrinsic frame, the harmonic transition operator has the expression:

\begin{equation}
T_{2\mu}=q_h\beta \left(\cos \gamma D^2_{M0}+\frac{\sin \gamma}{\sqrt{2}}
(D^2_{M2}+D^2_{M,-2})\right)\equiv q_h\beta{\cal T}_{2\mu}.
\end{equation}
The reduced matrix elements of the transition operator between the states described in the previous section are calculated in Appendix A. The reduced probability for the transition $|nvpIM\rangle \to |n'v'p'I'M'\rangle$ is obtained by squaring the corresponding reduced matrix element of the transition operator. As shown in Ref. \cite{Rad07} the final analytical expressions are very simple.

\section{Numerical results}
\label{sec:level4}
\renewcommand{\theequation}{4.\arabic{equation}}
\setcounter{equation}{0}

The equations derived in the semi-classical framework (\ref{En01}) for the energies of the states $0^+$  and $2^+$ were used to fit by a least square procedure the data for several nuclei: $^{152,154}$Gd, $^{162}$Dy, $^{168}$Er, $^{176}$Hf, $^{180,184}$W, $^{190}$Os. The parameters, $A, B, {\cal C}, F$ yielded by the fitting procedure are listed in Table I. The fitting procedure provides also the set of quantum numbers $\{n_k\}_k$ associated to the states specified by the ordering index $k$. Of course the values of $n$ provided by the equations expressing the condition that the $\chi^2$ value is minimum are not integers. We assigned to a given energy level $k$ the integer which is closest to $n_k$ yielded by the least square equations.     

The boson description provides the expressions (\ref{jeq0}) for energies of the states $0^+$, while 
for the states $2^+$, Eq.(\ref{jeq2}) is determining the energies. These equations define four sets of energies
which are depending on five parameters: $A, B, C, F, \gamma $. These parameters together with the quantum number N are to be fixed by a least square procedure.
For comparison we performed the fitting procedure for the same nuclei considered in the semi-classical approach.
Amazingly, both  procedures lead to a cubic expression in $n$ and $N$ respectively, although the two quantum numbers have different significance.
Indeed, the quantum number $n$ represents the number of the plane oscillator quanta associated to the intrinsic degrees of freedom $q_1, q_2$ related to the nuclear deformations $\beta, \gamma$. On the other hand the quantum number $N$ is the number of the quadrupole bosons which are describing the system in the laboratory frame.

Note that in principle both the boson number and seniority could be obtained by solving the least square equations but the procedure would be quite tedious. For the sake of simplicity we kept only the boson number as variable to be determined and chose the lowest seniorities. The reason is that for these seniority values the energy equation has a similar structure as in the semiclassical case. For the states $0^+$ we started with the $v=0$ expressions and tried to describe all energies as corresponding to $v=0$. The result was that a set of calculated energies exhibit large deviation from the experimental data. These states were considered to have v=3. In the next step both expressions, corresponding to  v=0 and v=3, have been used with the
assignments determined before, and new least square equations have been written for the five parameters and the $N_s$( number of states) values of $N$ (boson number).     
  
The results of the fitting procedures concerning the structure coefficients mentioned above are collected in Table I .
Inserting the fixed coefficients in the equations defining the energies, one obtains two sets of energies for classical description and four sets for the exact treatment. The sets of energies are plotted as functions of $n$ and $N$ respectively in Figs.1-16.

The four sets of energies, $ E_{N,0,0},E_{N,3,0}, E_{N,1,2}, E_{N,2,2}$ with the restrictions for $N$ mentioned above (see Eqs. \ref{jeq0} and \ref{jeq2}), are represented in the panels b),  c) of the left figures and panels  b),  c) of the right figures, respectively. In the panels a) of the left and right figures the semi-classical energies for the states $0^+$ and $2^+$ are given.  The full line curve is the energy as function of $N$, with $N$ considered as a continuous variable. The  integer number which lies closest to the experimental data is the assigned quantum number $N$. We remark that the agreement with the experimental data is quite good for both semi-classical and the exact eigenvalues.
The remarkable feature of our approach is that by  compact formulas we obtain a realistic description
of a large number of excitation energies, despite the fact that the number of the fitting parameters is relatively small.

\begin{table}
\begin{tabular}{|c|cccc|ccccc|}
\hline
            &\multicolumn{4}{c|}{Semi-classical method}&\multicolumn{5}{c|}{Exact solution}\\
	                                        \cline{2-10}
&$\hskip0.5cm$ A$ \hskip0.5cm$ &$ \hskip0.5cm $ B$ \hskip0.5cm$ &$ \hskip0.5cm$  C &$ \hskip0.5cm$  F$ \hskip0.5cm$  &
$ \hskip0.5cm $ A $ \hskip0.5cm $     &$ \hskip0.5cm$   B $ \hskip0.5cm $   &$ \hskip0.5cm $ ${\cal C}$ $ \hskip0.5cm$  & $ \hskip0.5cm $  F$ \hskip0.5cm$ &$ \hskip0.5cm$   $\gamma$   \\
\hline
$^{152}$Gd  &548.789  &-21.281  &-141.8  &1.135  &  343.827 &-20.827  &-8.0   &  1.135    & 161.443  \\
$^{154}$Gd  &294.746  &-6.726   &-40.    &0.21   &  234.694 &-6.642   &-40.   &0.21       &  31.978                 \\
$^{162}$Dy  &1063.4   &-93.     &-470.   &11.    &321.      &-88.6    &120    &11.        &621.2                               \\
$^{168}$Er  &394.2    &-10.4    &-280.   &0.3865 &83.2319   &-10.2454 &24.    &0.3865     &299.8638                  \\
$^{176}$Hf  &1140.6   &-111.787 &-700.   &15.145 &114.329   &-105.729 &32.    &15.145     &841.961              \\
$^{180}$W   &1030.37  &-92.32   &-656.   &11.45 &104.28    &-87.74   &-40    &11.45      &747.06                    \\
$^{184}$W   &781.966  &-50.0314 &-523.   &4.3282 &111.469   &-48.3001 &48.    &4.3282     &596.225             \\
$^{190}$Os  &856.71   &-52.904  &-567.372&4.51   &133.332   &-50.472  &-324.  &4.51       &504.188\\
\hline
\end{tabular}
\caption{The structure coefficients yielded by the fitting procedure applied to the semi-classical expressions for the $0^+$ and $2^+$ energies are given in the first four columns while the results for  those involved in the exact eigenvalue expressions are given in the last five columns. All coefficients are given in units of $keV$.}
\end{table}

\begin{figure}[ht!]
\begin{minipage}[t]{8cm}
\epsfysize=8cm
\centerline{\epsfbox{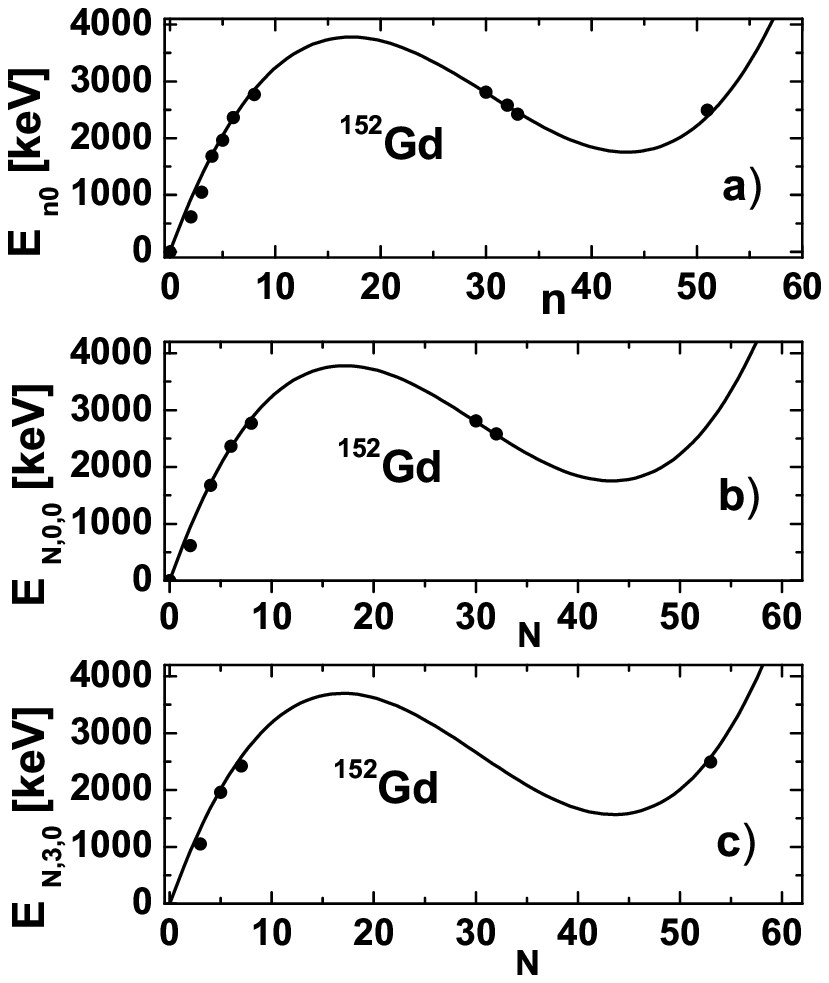}}
\caption{ Excitation energies of the  $J^{\pi}=0^+$ states, in $^{152}$Gd,
described semi-classically, panel a), and by eigenvalues of the model Hamiltonian, corresponding to the seniority
v=0, panel b), and v=3, panel c),
 are compared with the experimental data. }
\end{minipage}
\hspace{\fill}
\begin{minipage}[t]{8cm}
\epsfysize=8cm
\centerline{\epsfbox{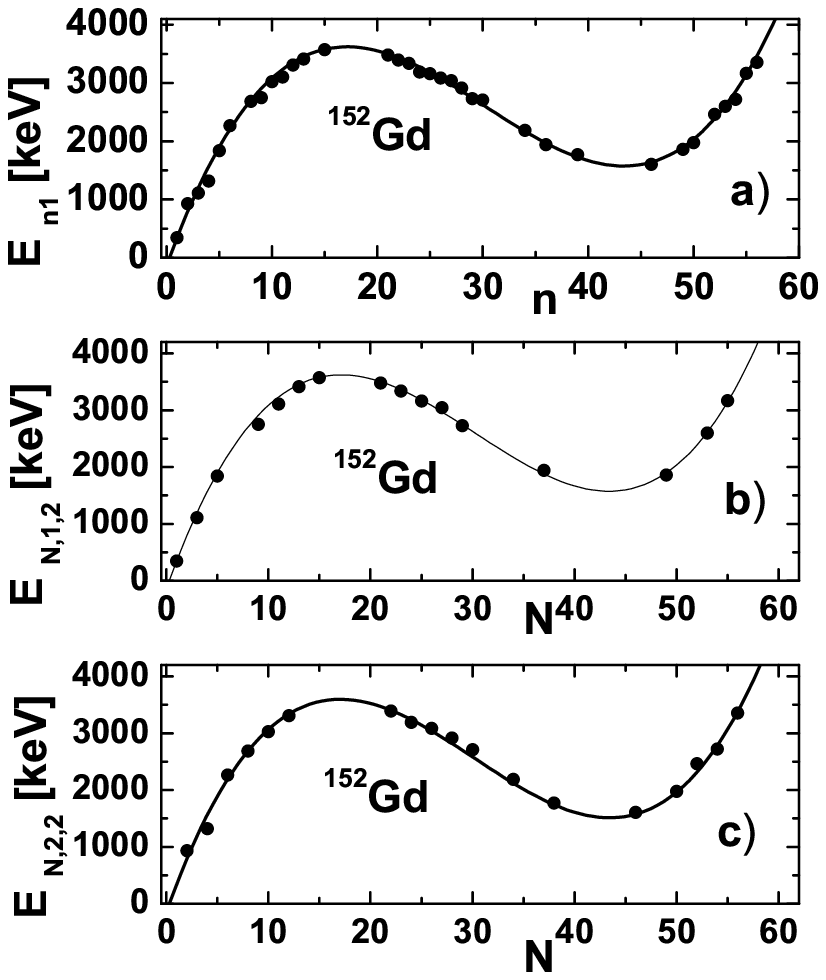}}
\caption{ Excitation energies of the $J^{\pi}=2^+$ states, described semi-classically, panel a), and by eigenvalues of the model Hamiltonian, corresponding to the seniority v=1, panel b) and v=2, panel c),
 are compared with the experimental data.}
\end{minipage}
\end{figure}

The states studied in this paper have been populated, by several groups, in experiments like $(p,t), (t,p), (D,D'), 
(n,\gamma), (n,n')$, Coulomb excitation.

For $^{152}$Gd, the energies of 11 states $0^+$ and 34 states $2^+$ are known from Ref.\cite{Rich,Agd}. The energies for $0^+$ states are smaller than $3000keV$ while the states $2^+$ lie below $3500keV$. In the case of $^{154}$Gd, 15 $0^+$ and 60 $2^+$ are known from Refs. \cite{Rich,Elb,Lon,Blo,Hami,Elz}.
The results for $^{152}$Gd are plotted in Figs. 1,2 while those for $^{154}$Gd are represented in Figs. 3, 4.

\begin{figure}[ht!]
\begin{minipage}[t]{8cm}
\epsfysize=8cm
\centerline{\epsfbox{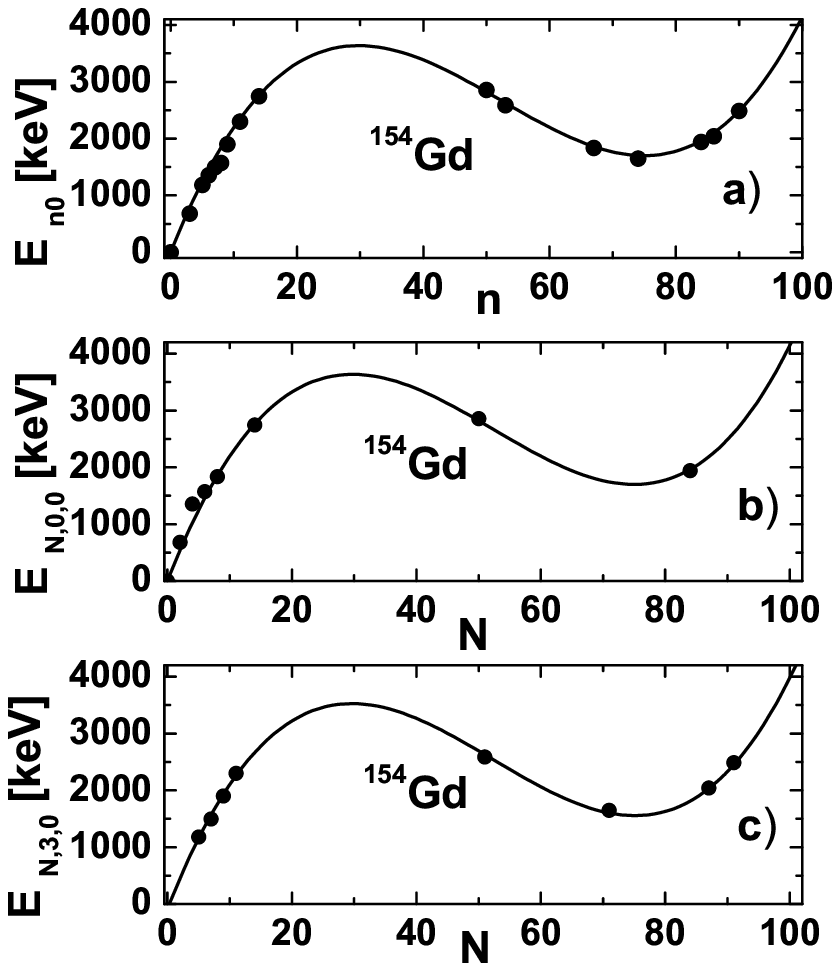}}
\caption{ The same as in Fig.1 but for $^{154}$Gd.}
\end{minipage}
\hspace{\fill}
\begin{minipage}[t]{8cm}
\epsfysize=8cm
\centerline{\epsfbox{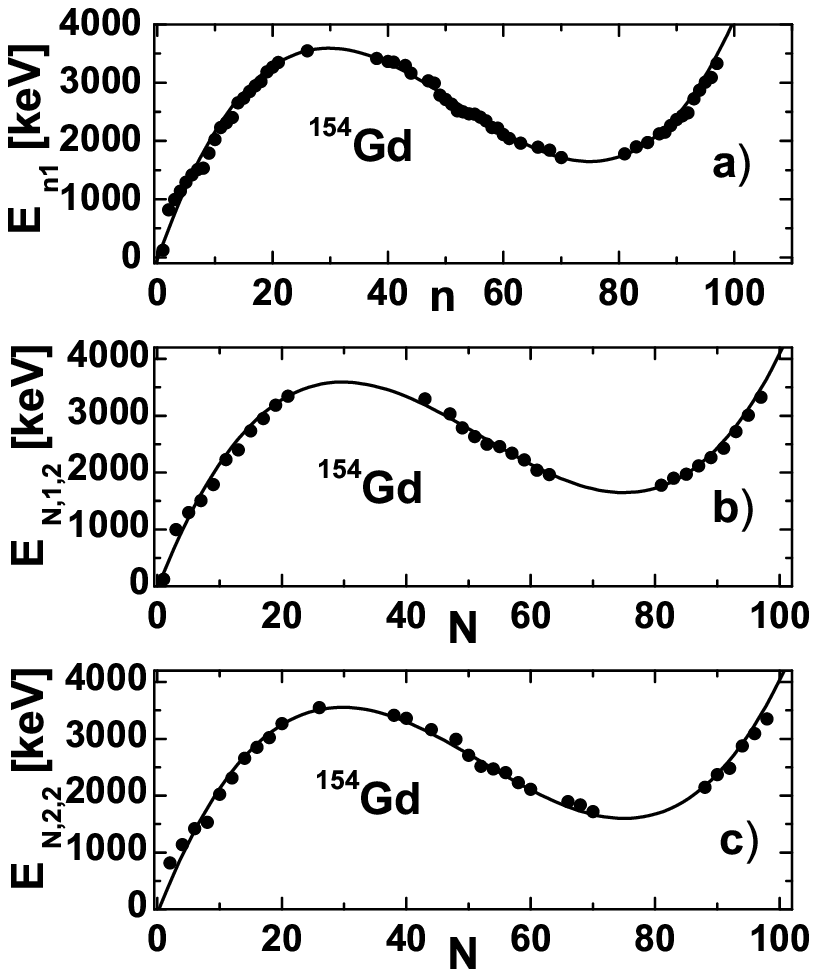}}
\caption{ The same as in Fig. 2 but for $^{154}$Gd.}
\end{minipage}
\end{figure}

Classical results can be interpreted in terms of quantized states of classical trajectories describing the motion in the potential V(q), defined by Eq.(\ref{Vdeq}). This potential has been plotted in Fig. 17 for $^{152}$Gd and $^{154}$Gd, respectively and in Fig.18 for $^{168}$Er, using alternatively the set of parameters provided by the semiclassical description and the exact treatment.

We recall the fact that along the isotopic chain of $Gd$ one records a transition from  spherical nuclei (the light ones) to  deformed like nuclei.
The first set of nuclei satisfy an $SU(5)$ symmetry while the second one an $SU(3)$ symmetry. The critical nucleus for this transition is considered to be $^{154}$Gd, which itself exhibits a distinct symmetry called $X(5)$ symmetry. This transition critical point is characterized by a specific value for the ratio $E_{4^+}/E_{2^+}$ and special features in the E2 properties of the ground as well as of the adjacent bands. The question is whether we find some fingerprints for this shape transition in the semiclassical description. This is in fact the reason we present here the potential energy corresponding to the two even isotopes of Gd.

Comparing the potential for the two Gd isotopes considered here, we note that the slope of $V(q)$ in its ascending part is higher for $^{152}$Gd than for $^{154}$Gd. In other words the first minimum is more flat for $^{154}$Gd than for $^{152}$Gd. The consequence is that the first $2^+$ state is lower in energy for $^{154}$Gd than for $^{152}$Gd. Since the transition for the state $4^+$ is felt less strongly than in the state $2^+$ \cite{RadFas}, the ratio of the two states energies is of course seriously affected. The second remark refers to the fact that the secondary minimum for $^{154}$Gd  is more deformed than that corresponding to $^{152}$Gd.

\begin{figure}[ht!]
\begin{minipage}[t]{8cm}
\epsfysize=8cm
\centerline{\epsfbox{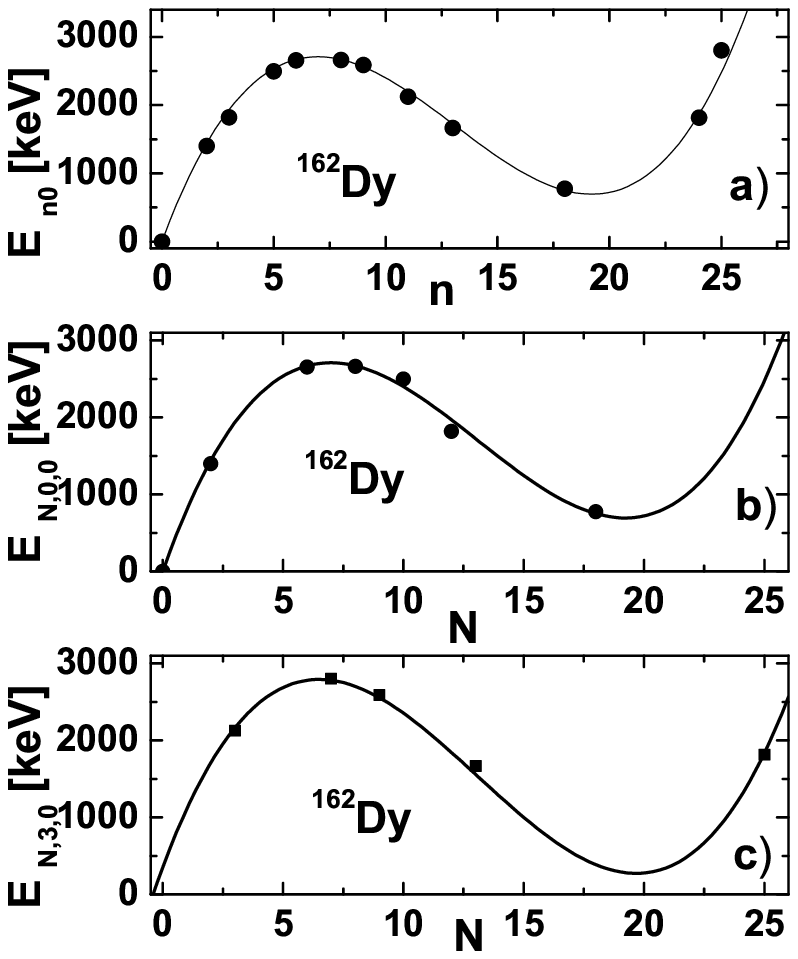}}
\caption{ The same as in Fig. 1 but for $^{162}$Dy.}
\end{minipage}
\hspace{\fill}
\begin{minipage}[t]{8cm}
\epsfysize=8cm
\centerline{\epsfbox{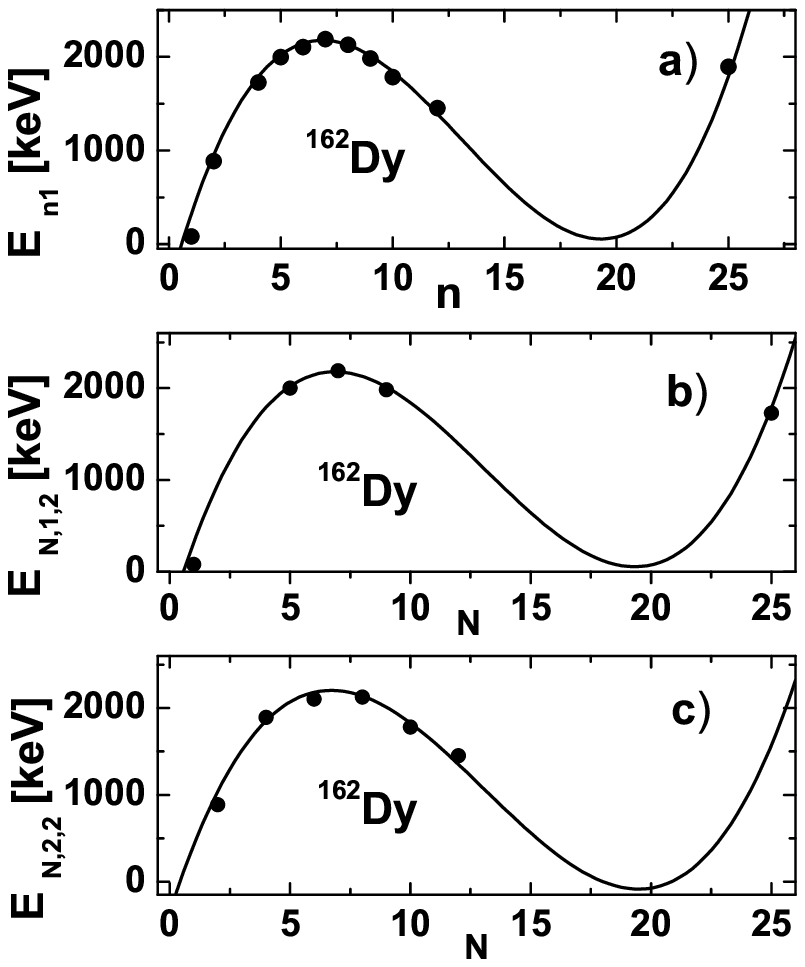}}
\caption{ The same as in Fig. 2 but for $^{162}$Dy.}
\end{minipage}
\end{figure}

In $^{162}$Dy have been observed 12 states $0^+$ and 11 states $2^+$ \cite{Rich,Helm}. The results of our calculations are compared with the corresponding experimental data in Figs.5, 6. We note that by contrast to the case of $Gd$' s isotopes here the energies are distributed around the maximum of the curves. Also we remark the low states density around the secondary minimum.

\begin{figure}[ht!]
\begin{minipage}[t]{8cm}
\epsfysize=8cm
\centerline{\epsfbox{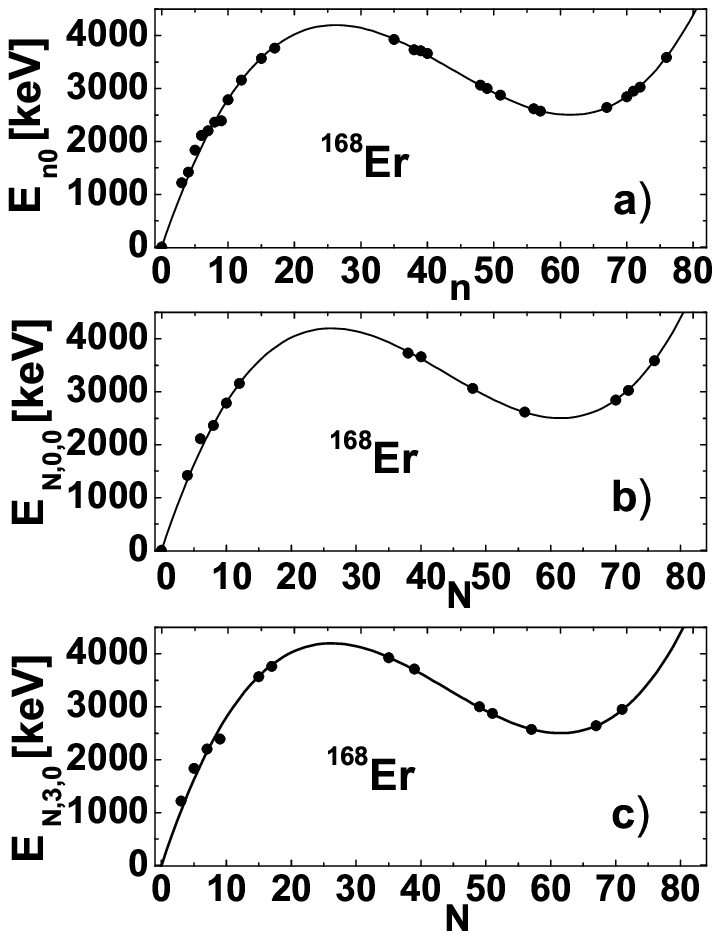}}
\caption{ The same as in Fig. 1 but for $^{168}$Er.}
\end{minipage}
\hspace{\fill}
\begin{minipage}[t]{8cm}
\hspace*{2cm}
\epsfysize=8cm
\centerline{\epsfbox{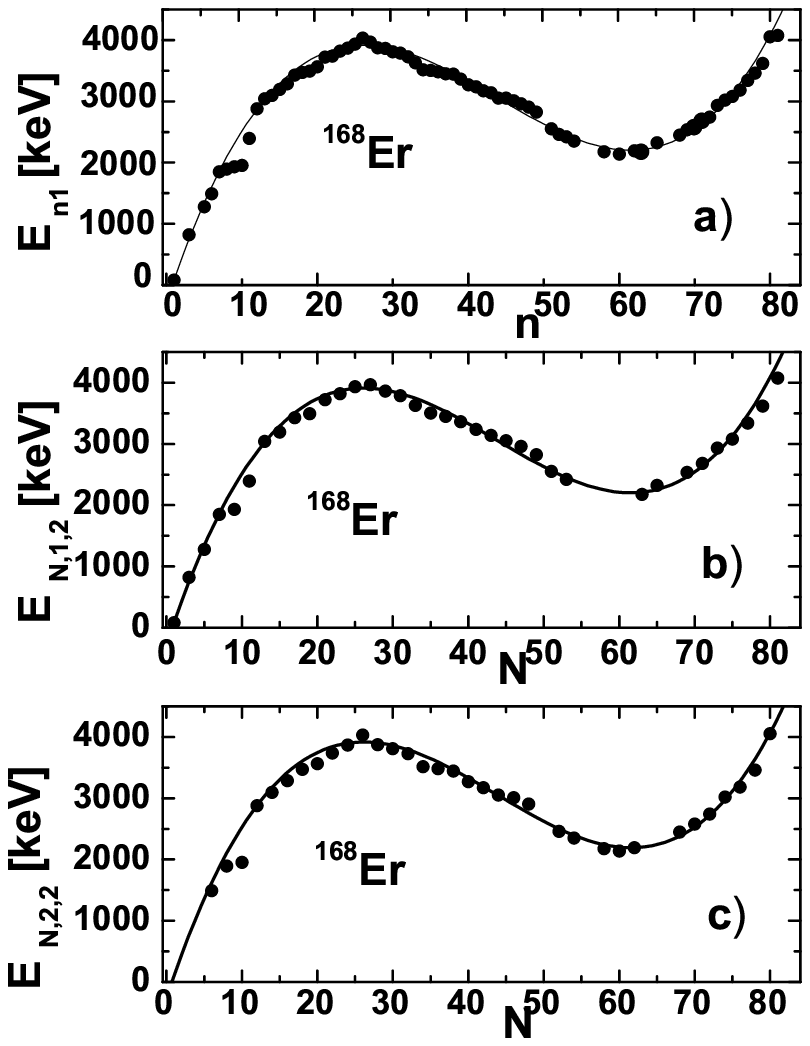}}
\caption{ The same as in Fig. 2 but for $^{168}$Er.}
\end{minipage}
\end{figure}
In $^{168}$Er, by a $(p,t)$ experiment, there have been identified 26 energy levels $0^+$ and 79 states $2^+$ \cite{Bucur,Bucur1}. 

As in any other theoretical model, the number of predicted energy levels in our model
is equal to the number of the considered basis states which, in
general,  is different from the number of the experimentally  identified
states. Clearly, in the plots shown here there are more predicted energy levels than experimental energies.

The predicted
energies which do not have correspondent data may require higher resolution or a different type of experiment. 
For example for $^{158}$Gd, the authors of Ref.\cite{Mey} found several new states through an $(p,t)$ experiment, that could not be seen by the previous $(n,n^{\prime})$ experiment \cite{Les}.

Concerning the predictive power of the present formalism it is worth mentioning an interesting story concerning the case of $^{168}$Er.
Indeed, after the publication of data  in Ref. \cite{Bucur}, where only 67 levels $2^+$ have been reported, and shortly after we provided a phenomenological
interpretation  in Ref.\cite{Rad07}, Bucurescu and his collaborators analyzed more carefully the data and found another 12 energy levels with angular momentum 2 and positive parity \cite{Bucur1}.  Of course, this was a challenge for us since explaining the new data is indeed a severe test for the proposed theoretical description. These 12 new levels are also considered here, keeping the fitted parameters from Ref. \cite{Rad07} unchanged.

\begin{table}
\begin{tabular}{|c|ccc|}
\hline
New energies[keV]&$\hskip0.2cm$N$\hskip0.2cm$   &$\hskip0.2cm$  v$\hskip0.2cm$  & Old energies[keV]\\
\hline
(2174.0)         & 63  & 1 &                  \\
2580.4           & 70  & 2 &                   \\
2683.2           & 71  & 1 &                   \\
2969.3           & 47  & 1 &2.961.2\\
3391.1           & 39  & 1 &3361.9\\
3418.2           & 17  & 1 &3429.2\\
3794.1           & 31  & 1 &3789.5\\
3838.0           & 29  & 1 &3861.9\\
(3923.4)         & 25  & 1 &3933\\
(4009.6)         & 26  & 2 &4033.5     \\
4060.7           & 80  & 2 &4055\\
4069.2           & 81  & 1 &4075.6\\
\hline
\end{tabular}
\caption{ Twelve  data for the newly identified  $2^+$ states are given on the left column. The first three are falling on the
graph representing the data from Ref.\cite{Bucur} and are interpreted as v=1 and v=2 states of large N. The remaining energies from the first column lie closely to the old energies from Ref.\cite{Bucur} described as low seniority states.}
\end{table}
The new data for the $2^+$ energies are given on the first column of Table II. The first three values fill the vacancies in the curves of Ref. \cite{Rad07} and are presented here in Fig.8. The assigned quantum numbers $N,v$ are those given in the second column of Table II. The remaining data lie very
closely to the data which are already represented in Fig. 8. The later are given also in the fourth column of Table II, together with the correspondingly assigned $N$ and $v$ values. The quasi-degeneracy for the energy levels between 2.9 and 4.1 MeV, shown in Table II, may suggest that a symmetry exists.
However, as mentioned already before, the set of states $|Nv\alpha JM>$  \cite{Ghe,Apo} does not comprise any degeneracy for $J=2$.
This feature led us to the conclusion that the new energies from the first column might correspond to  $(N,v)$ values which are different than  those given on  columns 2 and 3. This suspicion is based on the nonlinear character of the equation in $N$ and $v$
\begin{equation}
E_{N,v,J}=\cal{E}.
\end{equation}
for a given value of $\cal{E}$. Keeping the same parameters as before, we obtained the theoretical values for energies given in Table III.
\begin{table}
\begin{tabular}{|c|ccc|}
\hline
New energies[keV]&Theory&$\hskip0.2cm$  N $\hskip0.2cm$  &$\hskip0.2cm$  v$\hskip0.2cm$\\
\hline
2969.3           &2961.4& 45  & 5\\
3391.1           & 3408.4&39  & 7\\
3418.2           & 3426.8&14  & 10\\
3794.1           &3783.7& 17  & 11\\
3838.0           & 3833.1&80  & 8\\
(3923.4)         &3921.9& 28  & 4\\
(4009.6)         &4007.5& 26  & 8 \\
4060.7           &4055.6& 26  & 10 \\
4069.2           &4068.8  &27 &11\\
\hline
\end{tabular}
\caption{The excitation energies for the newly identified $2^+$ states, first column, are compared with the predictions of
the energy expression (2.19), (2.20), given in second column, for $N$ and $v$ from third and fourth column, respectively.}
\end{table}
Concluding, the compact formula given by Eq.(2.20) may describe a large amount of data despite the fact that only few parameters are involved. Most of the 
data are described as low seniority states but for $^{168}$Er, there are also energy levels which correspond to high seniority states.

Finally, we remark that the least square procedure yields for the first excited $0^+$ state in $^{168}$Er, a value for the boson number equal to three. On the other hand in Ref.\cite{War}, by means of a $(n,\gamma)$ reaction a complete scheme of levels has been produced for $J< 6$ and 
$E_x < 2$ MeV. The result is that there is no state $0^+$ with an energy smaller than 1.217 MeV. Thus, it is an open question which deserves further consideration, whether there are specific selection rules which prevent the population of the predicted $N=2, v=0$ state by the experiments mentioned above.

\begin{figure}[ht!]
\begin{minipage}[t]{8cm}
\epsfysize=8cm
\centerline{\epsfbox{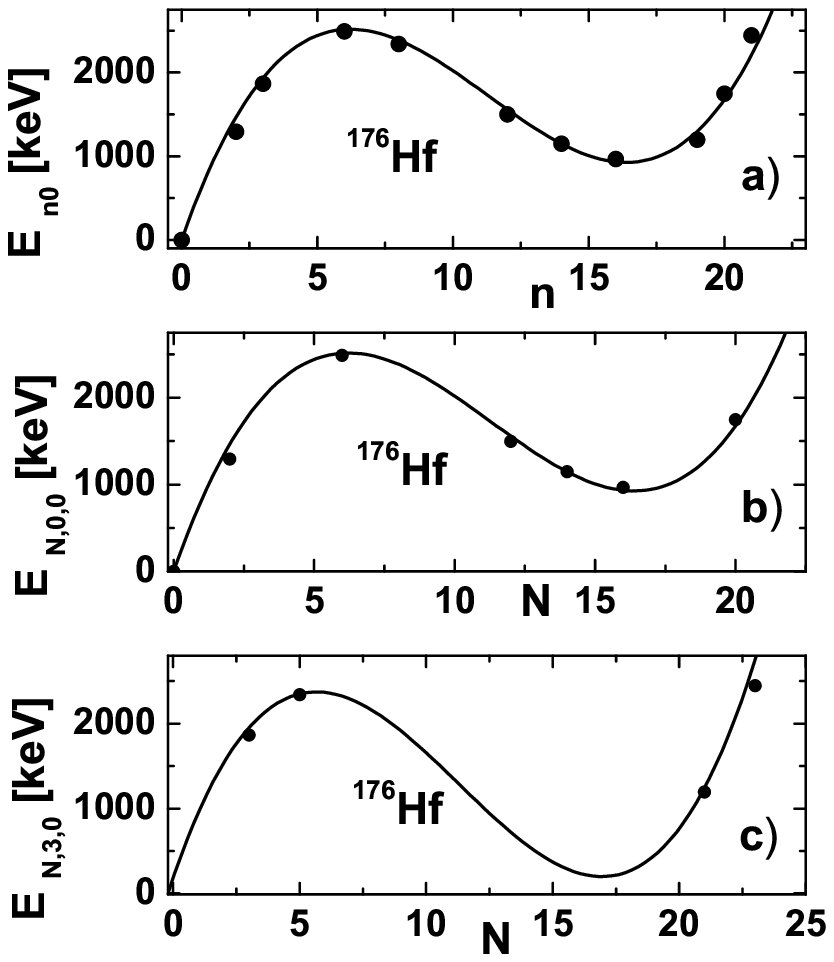}}
\caption{ The same as in Fig. 1, but for $^{176}$Hf.}
\end{minipage}
\hspace{\fill}
\begin{minipage}[t]{8cm}
\epsfysize=8cm
\centerline{\epsfbox{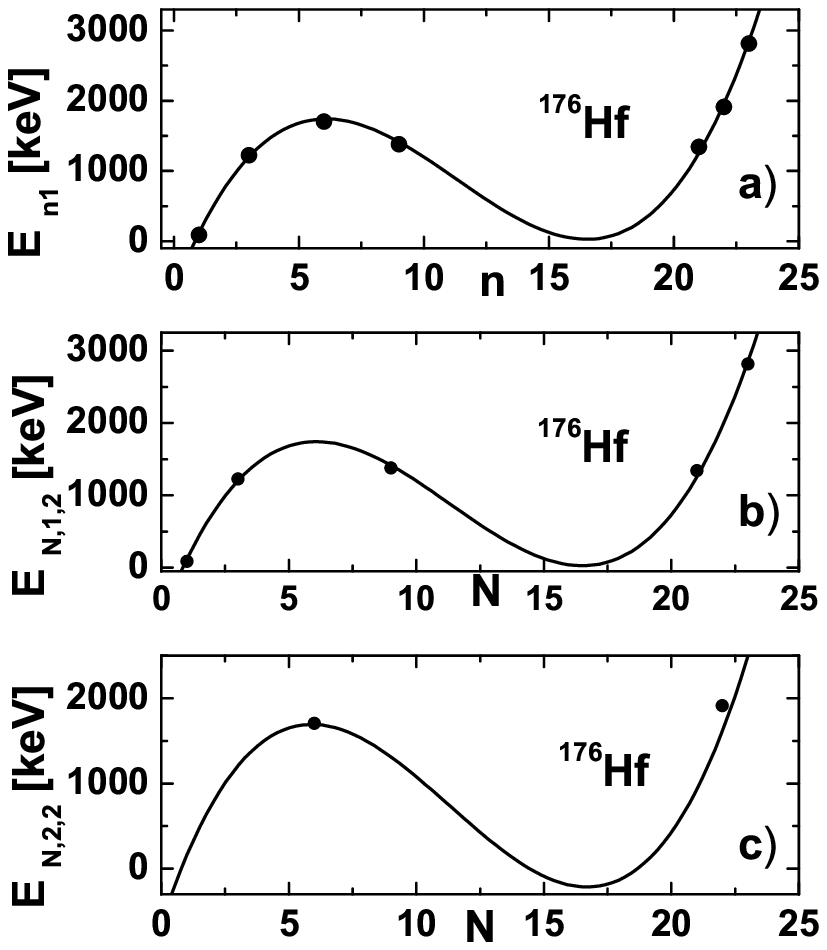}}
\caption{ The same as in Fig. 2, but for $^{176}$Hf.}
\end{minipage}
\end{figure}

In $^{176}$Hf, eleven $0^+$ and seven $2^+$ states are available \cite{Rich,Brow}. The theoretical results are plotted in Figs. 9,10 together with the experimental data. From there one notices that only few states of seniority 2 (1) and three (2) are found.

\begin{figure}[ht!]
\begin{minipage}[t]{8cm}
\epsfysize=8cm
\centerline{\epsfbox{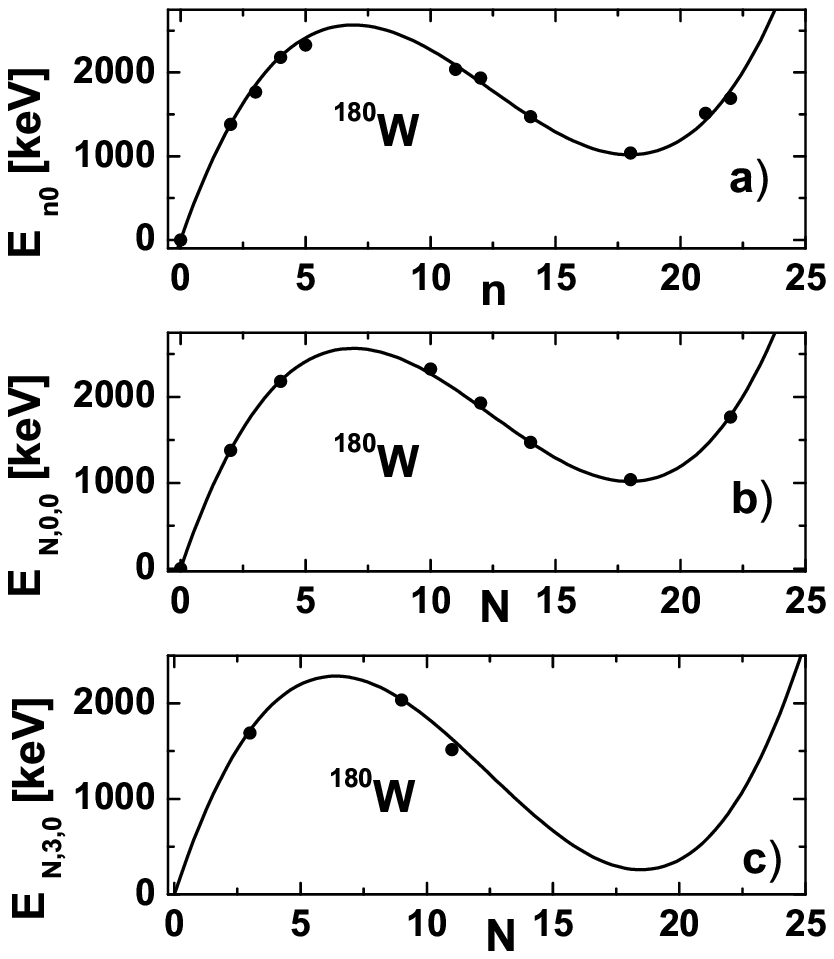}}
\caption{Fig. 11. The same as in Fig. 1, but for $^{180}$W.}
\end{minipage}
\hspace{\fill}
\begin{minipage}[t]{8cm}
\epsfysize=8cm
\centerline{\epsfbox{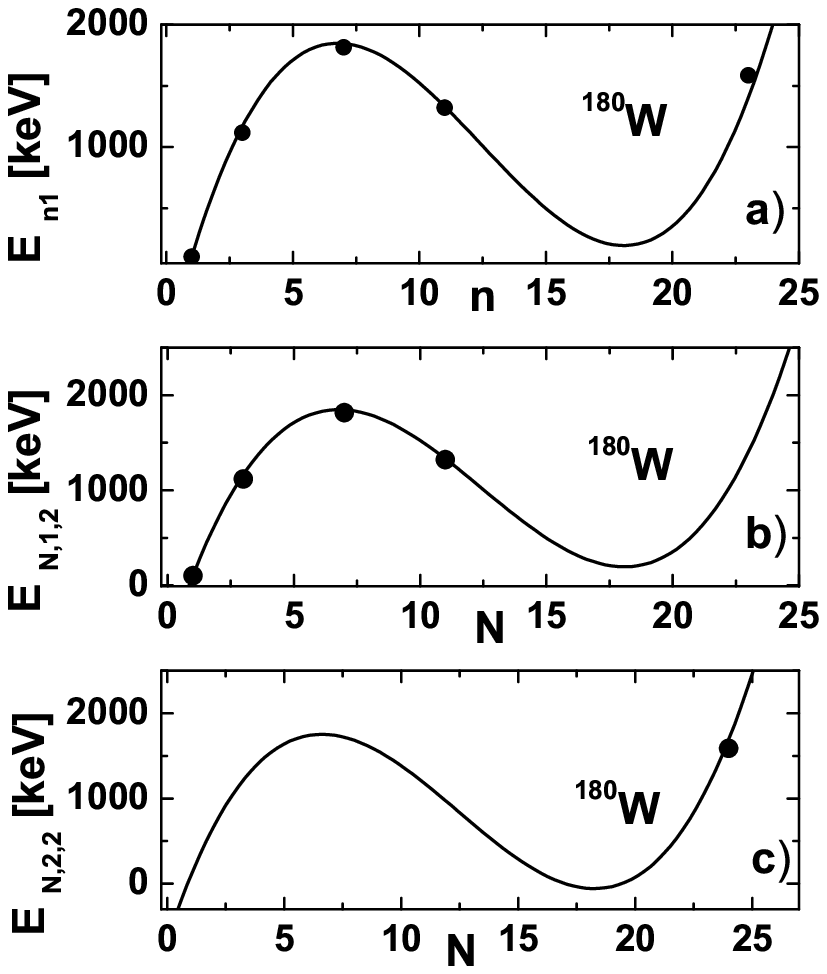}}
\caption{ The same as in Fig. 2, but for $^{180}$W.}
\end{minipage}
\end{figure}
For $^{180}$W one knows, from Ref.\cite{Rich,Wu}, the energies of eleven $0^+$ and five $2^+$.
Here we assigned the seniority 2 only to one state $2^+$ and seniority 3 only to two states $0^+$.

\begin{figure}[ht!]
\begin{minipage}[t]{8cm}
\epsfysize=8cm
\centerline{\epsfbox{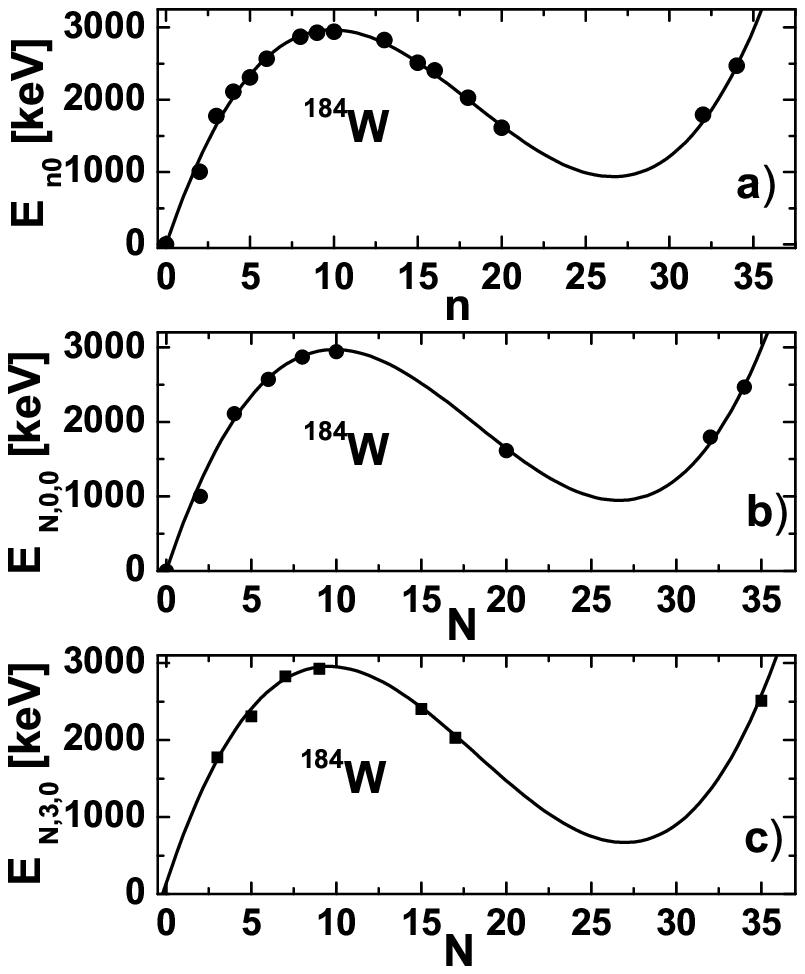}}
\caption{Fig. 13. The same as in Fig.1, but for $^{184}$W.}
\end{minipage}
\hspace{\fill}
\begin{minipage}[t]{8cm}
\epsfysize=8cm
\centerline{\epsfbox{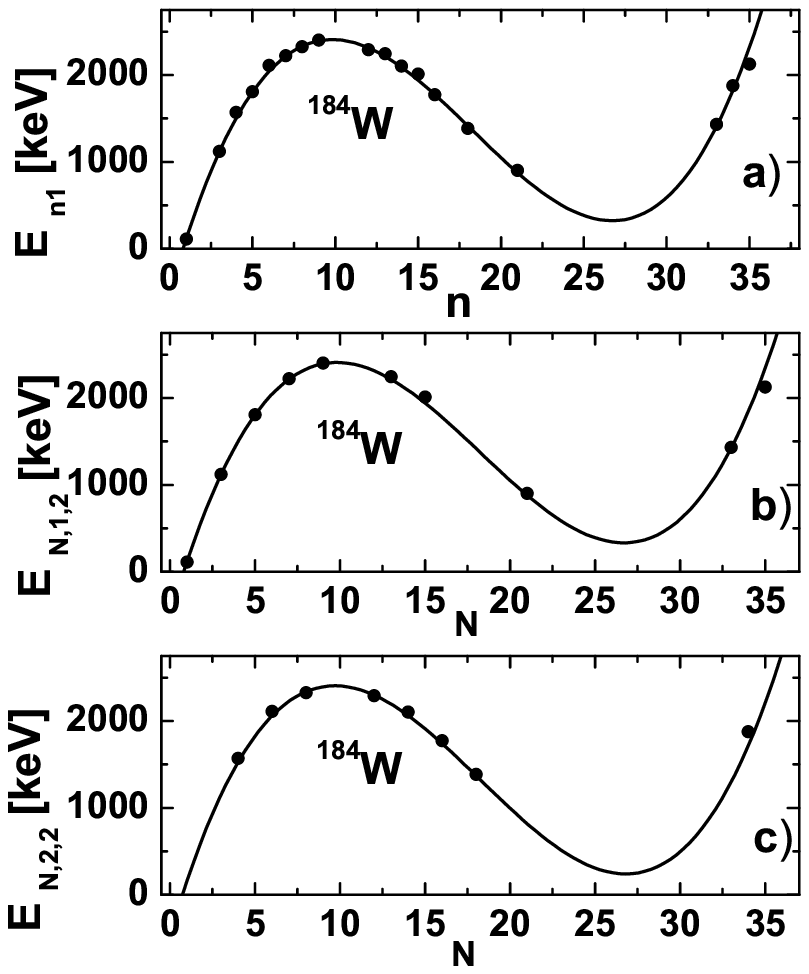}}
\caption{ The same as in Fig. 2, but for $^{184}$W.}
\end{minipage}
\end{figure}
For $^{184}$W more states are experimentally known. Indeed, in Refs. \cite{Rich,Fire} the energies of sixteen $0^+$ and eighteen $2^+$ have been reported.

\begin{figure}[ht!]
\begin{minipage}[t]{8cm}
\epsfysize=8cm
\centerline{\epsfbox{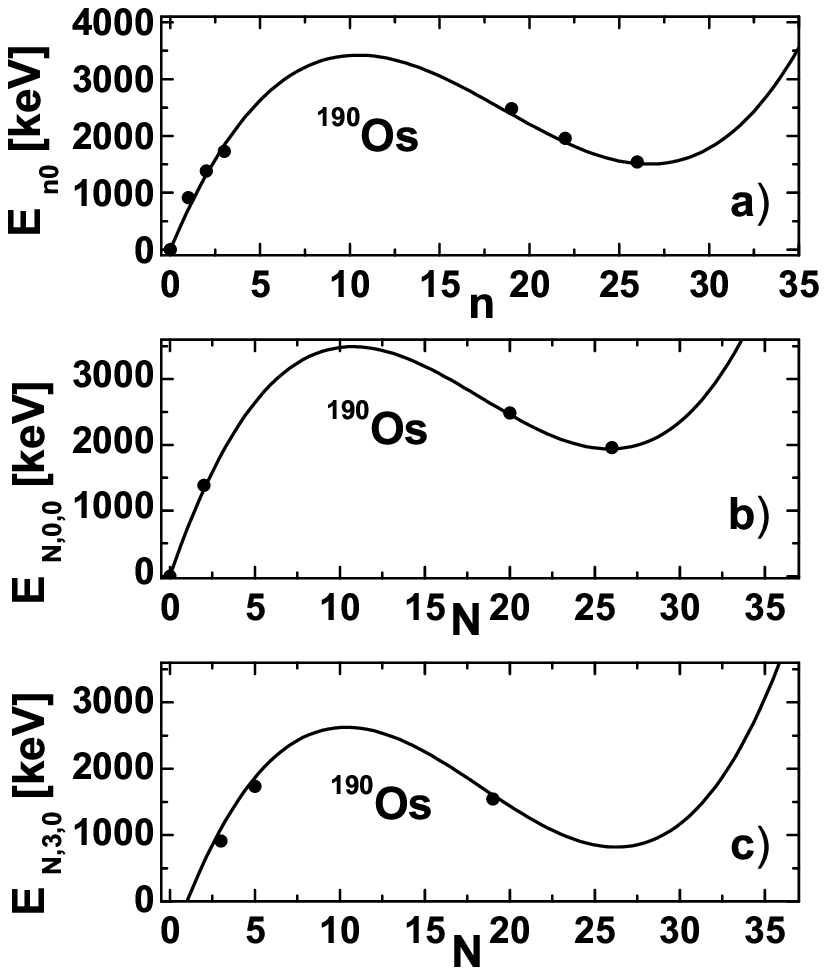}}
\caption{ The same as in Fig.1, but for $^{190}$Os.}
\end{minipage}
\hspace{\fill}
\begin{minipage}[t]{8cm}
\epsfysize=8cm
\centerline{\epsfbox{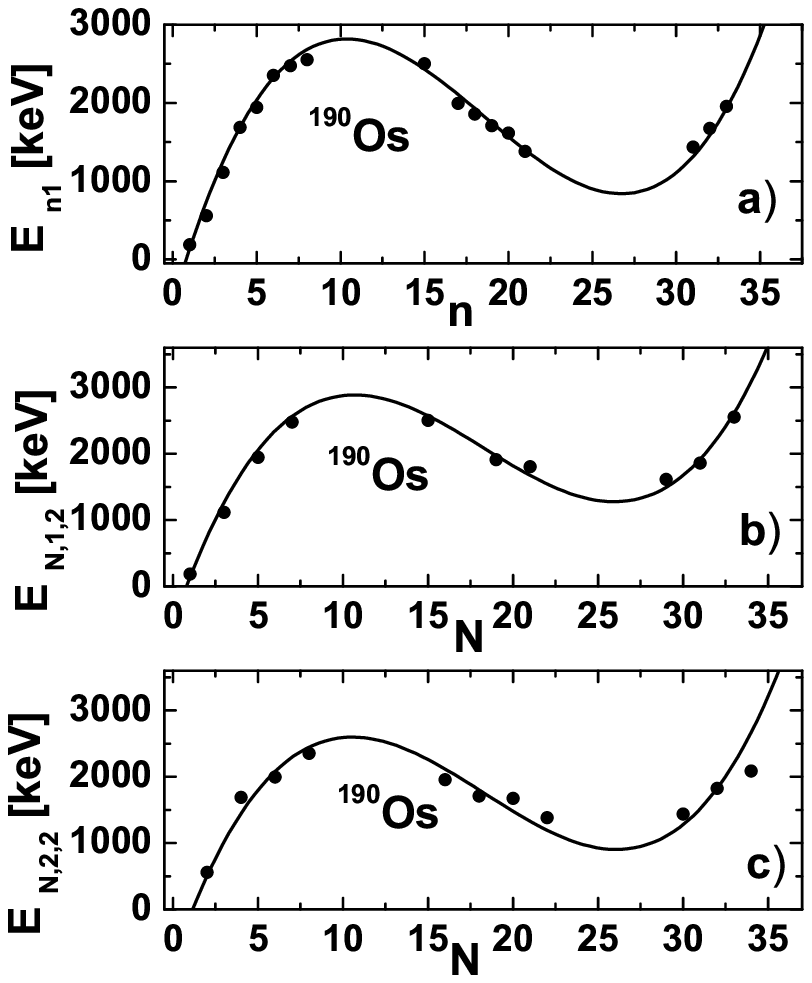}}
\caption{ The same as in Fig. 2, but for $^{190}$Os.}
\end{minipage}
\end{figure}
The last nucleus investigated is $^{190}$Os for which we know seven energy levels $0^+$ and seventeen $2^+$.
Experimental data are those from Ref.\cite{Rich,Fire}.

\begin{figure}[ht!]
\begin{minipage}[t]{8cm}
\epsfysize=8cm
\centerline{\epsfbox{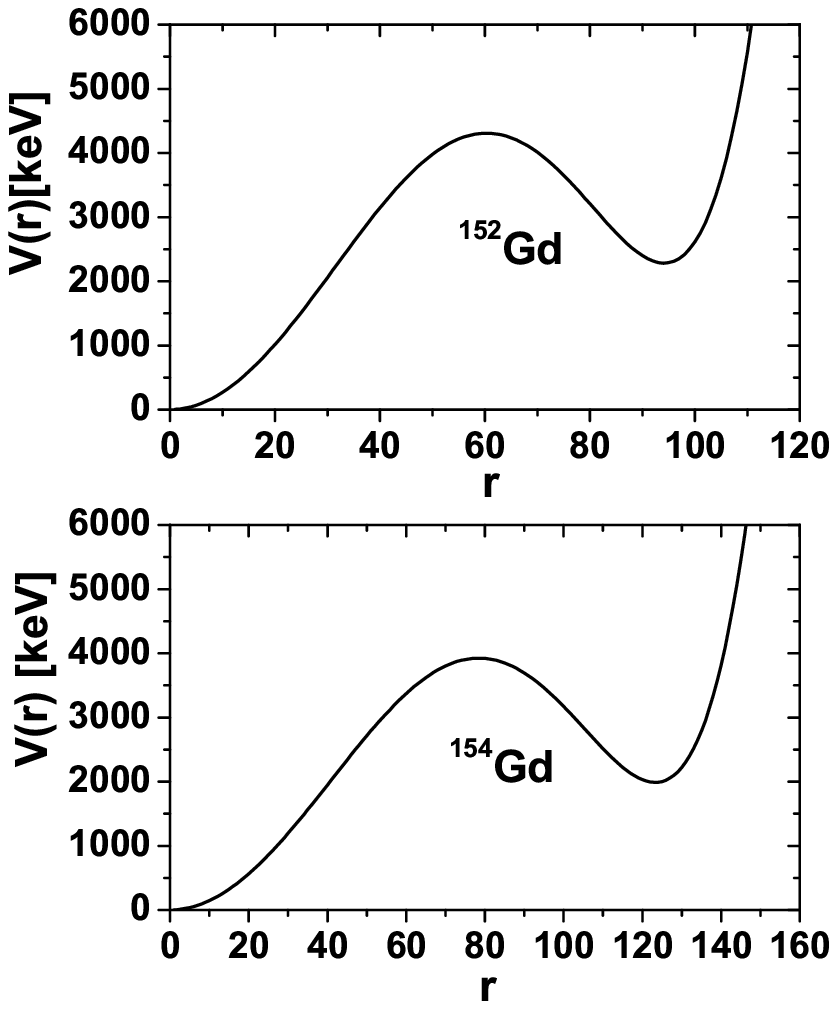}}
\caption{The classical potentials (2.7) corresponding to the sets of parameters specified in Table I for $^{152}$Gd (upper panel) and $^{154}$Gd (lower panel) provided by the semi-classical description are plotted as functions of the polar coordinate $r(=\sqrt(q^2_1+q^2_2))$.}
\end{minipage}
\hspace{\fill}
\begin{minipage}[t]{8cm}
\epsfysize=8cm
\centerline{\epsfbox{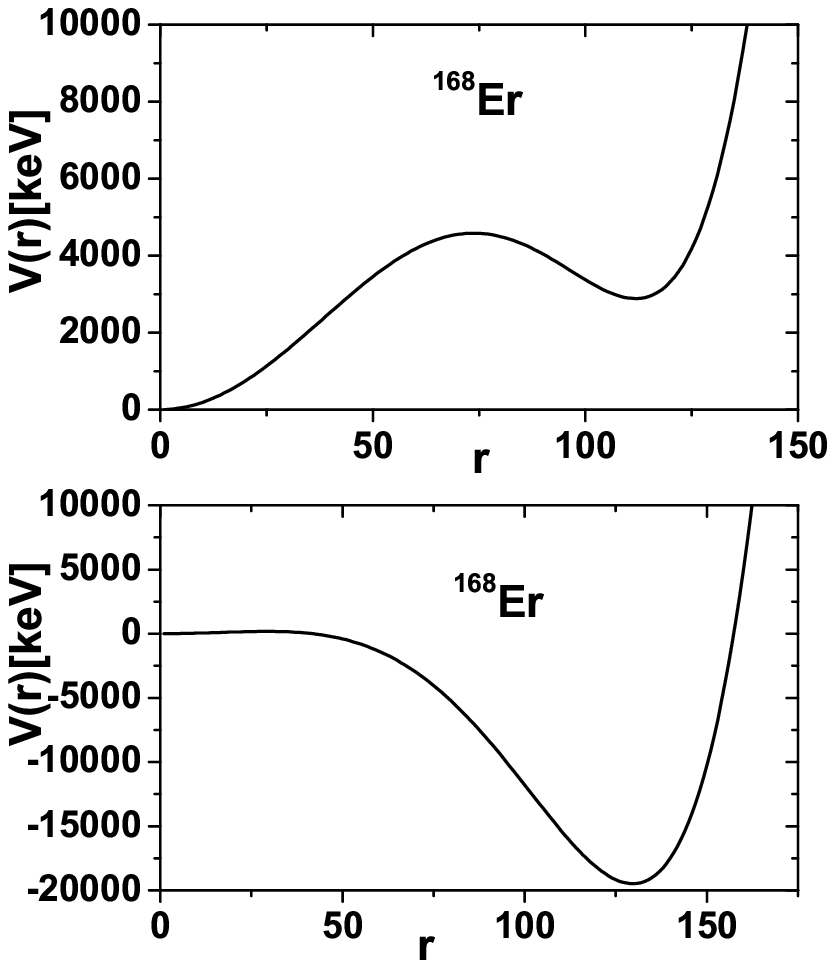}}
\caption{The classical potentials (2.7) corresponding to the sets of parameters specified in Table I for $^{168}$Er, provided by the semi-classical (upper panel) and exact descriptions (lower panel) are plotted as functions of the polar coordinate $r(=\sqrt(q^2_1+q^2_2))$.}
\end{minipage}
\end{figure}

\clearpage

\section{Conclusion}
\label{sec:level 5}
\renewcommand{\theequation}{5.\arabic{equation}}
\setcounter{equation}{0}
In the previous sections, we proposed two  phenomenological descriptions of the excitation energies of the states $0^+$ and $2^+$ experimentally identified in several  even-even nuclei.
They correspond to two distinct ways of treating the same sixth-order quadrupole boson Hamiltonian. One is a semi-classical description while the second one uses the exact eigenvalues. While in the yrast band the highest seniority states are the best candidates for a realistic description, for the states of the same angular momentum,  the lowest seniority states are used for most  states. We found, however, that some of the states $2^+$ of $^{168}$Er are higher seniority states.
It is remarkable that both $0^+$ and $2^+$ states exhibit a cubic $n$ dependence. We know that such a behavior for energy in the yrast bands is determining a back-bending \cite{Rege} phenomenon for the moment of inertia as a function of the rotational frequency.
Here a back-bending  also shows up but the cause is different from that determining the bending in the moment of inertia in the yrast band.

The terms of the classical Hamiltonian which do not depend on momenta define the potential of the classical system. This has been plotted in Fig.17 for $^{152,154}$Gd and Fig. 18 for $^{168}$Er. For $Gd$ isotopes we used the parameters provided by the semi-classical treatment while for $^{168}$Er the two panels correspond to  two sets of parameters obtained by  classical and exact descriptions, respectively. From the upper panel of Fig. 18 we notice that some semi-classical states may accommodate the second well of the potential\cite{Rad3}. The boson description yields a similar spectrum as the semi-classical method but with different structure parameters, i.e. those from Table I corresponding to the columns with the exact solutions. The quoted parameters define a classical potential, given in the lower panel of Fig. 18, which is very different from the one used in the classical picture. The discrepancy is caused by the high anharmonicities involved. Actually, the two pictures, semi-classical and quantal, agree with each other only in the harmonic limit. Comparing the potentials for the two isotopes of $Gd$ one finds a qualitative explanation for the behavior of the ratio $E_{4^+}/E_{2^+}$ which suggests that $^{154}$Gd is a good candidate for the critical point in the shape phase transition which takes place in the chain of $Gd$ even isotopes.

It is worth noticing that  for a long time, theoretical works were focused on explaining the high spin states in the ground band, but not so much was done about bands aside the ground state band. Now we are confronted with a new situation. Indeed, to explain consistently very many excitation energies of states with low angular momenta is a real challenge for any theoretical approach. For example in $^{168}$Er a large number (105) of energy levels are experimentally known, 26 of spin 0 and 79 of spin 2. 

Using a sixth-order boson Hamiltonian we derived analytical formulas for the excitation energies of these states which involve a small number of parameters: four in the semi-classical treatment and five in the boson description. Both sets of formulas are describing quantitatively quite well the existent data. In order to draw a conclusion about how these coefficients depend on the atomic mass a richer systematics is necessary.

One may argue that for  many of the states considered here, the single particle degrees of freedom prevail. Actually we may share this opinion but, on the other hand, we think that the single particle behavior may be simulated by the anharmonicities involved in the present phenomenological model. Some of the considered states may have collective features. It is worth mentioning that the present model is able to account for these properties shown by a deformed nucleus such as $^{168}Er$ despite the fact that one uses a boson number conserving Hamiltonian. Our attempt is not singular in this respect. Indeed, this is one of the signatures of the interacting boson approximation
\cite{AriIache}
which is successful in describing rotational bands in non-spherical nuclei. Moreover, the Hamiltonian given by Eq. (1) with $F=0$,
has been  previously used to describe the yrast bands in transitional and deformed nuclei \cite{RadDreiz,DreizKlei}.
The results of the quoted papers show that some properties determined by the nuclear deformation can be described by
a suitable choice of the structure coefficients multiplying the anharmonic terms.
Certainly,  data  concerning the electromagnetic  transitions of these states are necessary in order to have an additional test and a more complete picture.

A very nice  test of the predictive power of our simple formulas was obtained by applying them  to the newly found data for $^{168}$Er, by keeping the numerical values for the structure coefficients as obtained in our previous calculations. We showed that the new data are surprisingly well described by the same parameters set.

Also, very simple formulas  for the B(E2) values characterizing the transitions between the states are derived within the two  approaches.
Note that the expressions for the semi-classical transition $|2(n-1)\rangle \to |0n\rangle$ and the boson transition $|n-1~102\rangle\to|n~000\rangle$
are identical.

Of course  the microscopic descriptions has the great merit of interpreting the data in terms of the single particle motion and may address some issues which are complementary to those accessible for phenomenological models. On the other hand in phenomenological pictures one may find a way to improve the microscopic description. Indeed, in our formalism we have seen that the sixth-order boson term is necessary in order to obtain a quantitative description of the data. On the other hand an RPA treatment of a Hamiltonian involving the mean field and a two body $Q.Q$ interaction, yields a quadratic boson term.  
Going beyond QRPA by a boson expansion procedure, higher order terms in bosons are obtainable. It is well known that the quadrupole two quasiparticle operator $A^{\dagger}_{2\mu}(a,b)$ can be expressed as an  odd powers expansion in bosons while the quadrupole quasiparticle density operator $B^{\dagger}_{2\mu}(a,b)$ as an even power. Therefore, ignoring the quasiparticle correlations due to the operators $A^{\dagger}_{2\mu}B^{\dagger}_{2\mu}, A^{\dagger}_{2\mu}B_{2\mu},
 A_{2\mu}B^{\dagger}_{2\mu}, A_{2\mu}B_{2\mu}$ and considering the first-order boson expansion for the remaining terms, one obtains a sixth-order boson Hamiltonian similar to the phenomenological Hamiltonian used in the present paper. Concluding one may assert that the phenomenological Hamiltonian has actually a microscopic counterpart. On the other hand our calculations suggest that going beyond QRPA the microscopic model QPM might describe in a better quantitative way the existent experimental data.

Recently the states $0^+$ have been considered by the Interacting Boson Approximation (IBA) approach within the phase transition context and several analytical results have been derived \cite{Bona}. Thus it was shown that in each of the symmetries $U(5)$, $O(6)$ and $SU(3)$ the energies of the state $0^+_n$ depend linearly on $n$ in the regime of large values for the quadrupole boson numbers. Several degeneracies of the states $0^+_k$ with the states $J^+_1$ have been pointed out for the critical values of the ordering parameters. In particular,  the degeneracy $E(0^+_2)\approx E(6^+_1)$ might be viewed as a hallmark of the $X(5)$ symmetry and moreover the ratio $E(6^+_1)/E(0^+_2)$ could play the role of the order parameter for the specific phase transition. For the symmetries characterizing the critical points of phase transitions, analytical expressions for the energies $E(0^+)$ have been derived. The energy $E(0^+_n)$ depends on $n$ as $n(n+x)$ with x depending on the considered phase transition. 

In order to make a fair comparison of the present results with those of Ref.\cite{Bona}, analytical results of $E(0^+)$ for any interaction strength of the IBA Hamiltonian would be desirable and moreover the same order boson Hamiltonian to be considered. By considering in the expressions \ref{jeq0} the situation when the six order term is missing, i.e. F=0, one obtains:

\begin{eqnarray}
E_{N,0,0}&=&BN(N+x),\;N=0,2,4,...
\label{jeq01}\\
E_{N,3,0}&=&BN(N+x)+\frac{9}{4}C, N=3,5,7,....\nonumber
\end{eqnarray}
with x=$(A+\gamma-\frac{3}{8}C)/B$.  
Thus, one may say that the results of Ref.\cite{Bona} may be recovered  by the present formalism in the limit of $F=0$. However, the six order term
determines an additional minimum in the classical potential energy and therefore a new phase of the nuclear system is expected. Moreover, the cubic $N$ dependence may account for the complex structure of the energy distribution for the states $0^+$ and $2^+$.

{\bf Acknowledgments.} A.A.R. wants to thank UCM-GRUPO SANTANDER for financial support of his visit at Complutense University of Madrid within the program of Distinguished Visitors, where part of this work has been performed. 
This work was supported  by the Romanian Ministry for Education and Research under the contract PNII, No. ID-33/2007.

\section{Appendix A}
\label{sec:level6}
\renewcommand{\theequation}{A.\arabic{equation}}
\setcounter{equation}{0}

Here we give the results for the matrix elements of the transition operators factors, which are needed for calculating the $B(E2)$ values. The matrix elements of $\beta$ are:
\begin{eqnarray}
\langle F_{n+1~1}|\beta |F_{n0}\rangle &=&\sqrt{\frac{n+5}{2}},\nonumber\\
\langle F_{n-1~1}|\beta |F_{n0}\rangle  &=&\sqrt{\frac{n}{2}},\nonumber\\
\langle F_{n+1~2}|\beta |F_{n3}\rangle  &=&\sqrt{\frac{n-1}{2}},\nonumber\\
\langle F_{n-1~2}|\beta |F_{n3} \rangle &=&\sqrt{\frac{n+6}{2}},\nonumber\\
\langle F_{n+1~2}|\beta |F_{n1} \rangle &=&\sqrt{\frac{n+6}{2}},\nonumber\\
\langle F_{n-1~2}|\beta |F_{n1} \rangle &=&\sqrt{\frac{n-1}{2}},\nonumber\\
\langle F_{n+1~1}|\beta |F_{n2} \rangle &=&\sqrt{\frac{n}{2}},\nonumber\\
\langle F_{n-1~1}|\beta |F_{n2} \rangle &=&\sqrt{\frac{n+5}{2}}.
\end{eqnarray}
With the convention of Rose for reduced matrix elements, one obtains:
\begin{eqnarray}
\langle {\cal G}_{102}||{\cal T}_2||{\cal G}_{000}\rangle &=&\frac{1}{\sqrt{5}},\nonumber\\
\langle {\cal G}_{202}||{\cal T}_2||{\cal G}_{310}\rangle &=&\frac{1}{\sqrt{15}},\nonumber\\
\langle {\cal G}_{202}||{\cal T}_2||{\cal G}_{102}\rangle &=&\sqrt{\frac{2}{7}},\nonumber\\
\langle {\cal G}_{102}||{\cal T}_2||{\cal G}_{202}\rangle &=&\sqrt{\frac{2}{7}}.\nonumber\\
\end{eqnarray}

\end{document}